\documentclass{article}
\usepackage{spconf,amsmath,graphicx,subcaption,float}
    
\usepackage{enumitem}
\setlist{nosep, leftmargin=14pt}

\usepackage{mwe} 

\title{SpineFM: Leveraging Foundation Models for Automatic Spine X-ray Segmentation}
\twoauthors
  {Samuel J. Simons}
	{Department of Engineering Science \\ University of Oxford\\ Oxford, UK}
  {Bart\l omiej W. Papie\.z}
	{Big Data Institute\\ University of Oxford\\ Oxford, UK}
\begin{document}
\ninept
\maketitle
\begin{abstract}
This paper introduces SpineFM, a novel pipeline that achieves state-of-the-art performance in the automatic segmentation and identification of vertebral bodies in cervical and lumbar spine radiographs.
SpineFM leverages the regular geometry of the spine, employing a novel inductive process to sequentially infer the location of each vertebra along the spinal column. Vertebrae are segmented using Medical-SAM-Adaptor, a robust foundation model that diverges from commonly used CNN-based models. We achieved outstanding results on two publicly available spine X-Ray datasets, with successful identification of 97.8\% and 99.6\% of annotated vertebrae, respectively. Of which, our segmentation reached an average Dice of 0.942 and 0.921, surpassing previous state-of-the-art methods.
\end{abstract}
\begin{keywords}
Foundation models, Spinal X-ray imaging, Vertebra segmentation
\end{keywords}
\section{Introduction}
\label{sec:intro}
In 2020, over 820 million individuals were estimated to suffer from neck and lower back pain due to spinal abnormalities or damage \cite{neck-pain-data,back-pain-data}. X-rays are frequently used to diagnose these conditions because of their availability and relatively low cost; however, manual annotation by clinicians remains both time-consuming and technically challenging, often due to patient-specific pathology variations and clinician fatigue \cite{shen_deep_2017,kundu2024spinal}. Artificial Intelligence (AI) has the potential to address these issues through automated analysis of images. However, a key foundation of many AI-based methods is accurate vertebra segmentation, making robust segmentation techniques essential for advancing the performance of the diagnostic tools.

Numerous studies have utilized convolutional architectures for vertebra segmentation. For instance, \cite{kim_automated_2021} proposed a two-step approach involving vertebra localization followed by segmentation, while \cite{VertXNet} introduced a rule-based ensemble model to extract vertebral bodies. Although these methods have shown promising results, they face significant limitations, including limited transferability to new datasets and high demands for data annotations. Recently, foundational models have emerged as a competitive alternative. These models leverage massive training datasets and architectures with hundreds of millions of parameters to develop a comprehensive understanding of the task, achieving competitive zero-shot performance without requiring extensive fine-tuning.

\begin{figure}[t]  
    \centering
    \begin{subfigure}[b]{0.15\textwidth}  
        \centering
        \includegraphics[width=\textwidth]{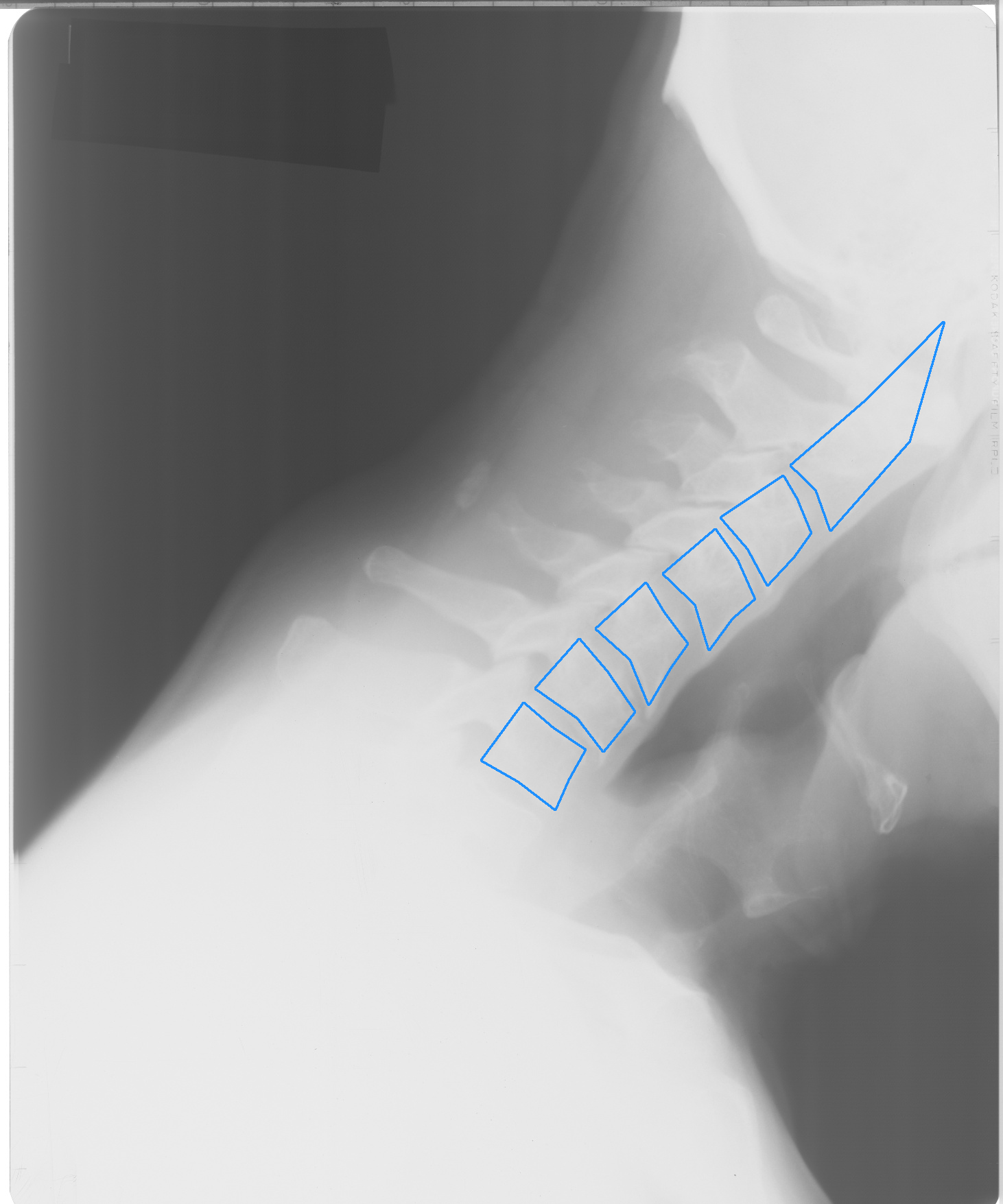}
        \label{fig:sub1}
    \end{subfigure}
    \hfill  
    \begin{subfigure}[b]{0.15\textwidth}  
        \centering
        \includegraphics[width=\textwidth]{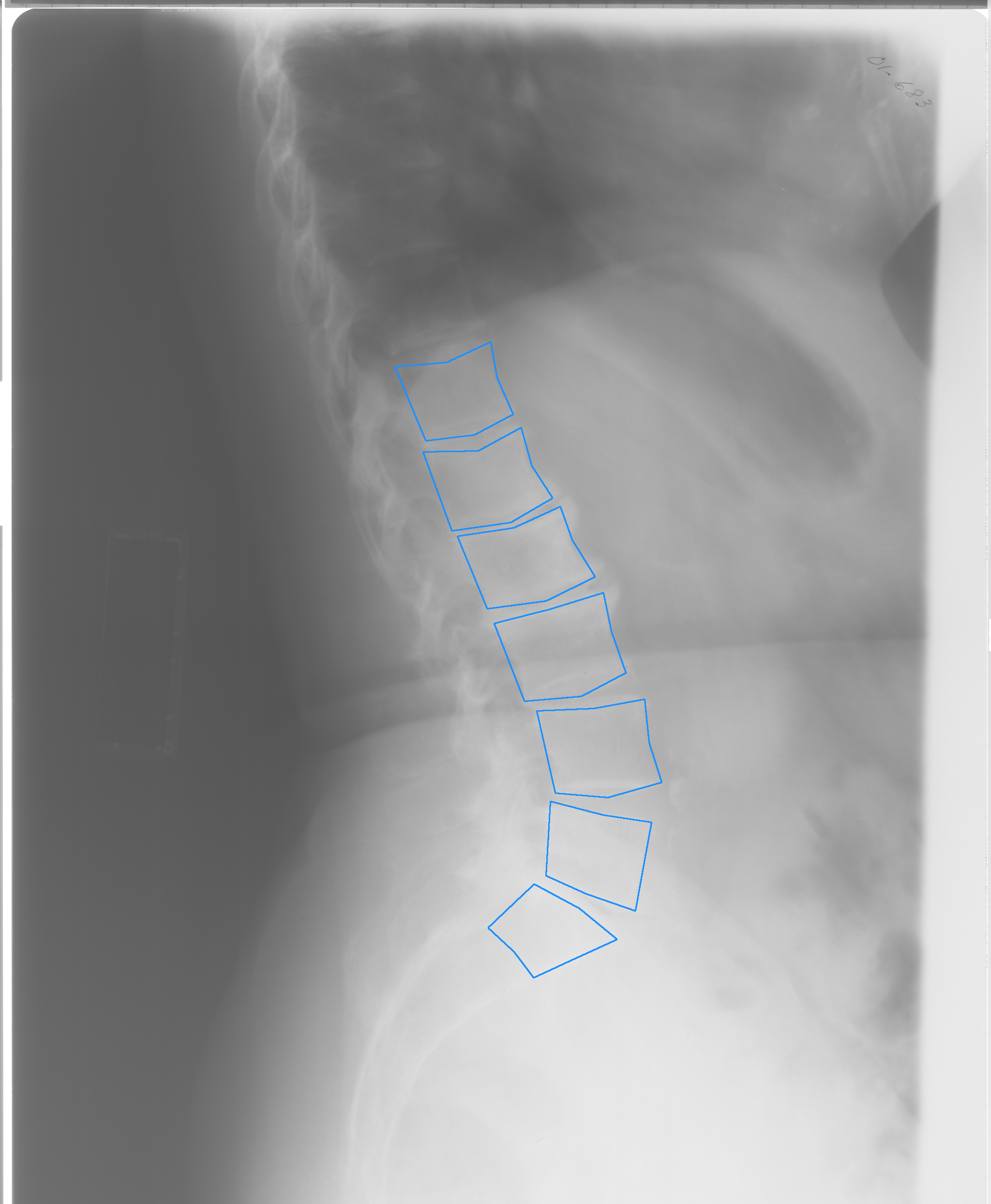}
        \label{fig:sub2}
    \end{subfigure}
    \hfill
    \begin{subfigure}[b]{0.155\textwidth}  
        \centering
        \includegraphics[width=\textwidth]{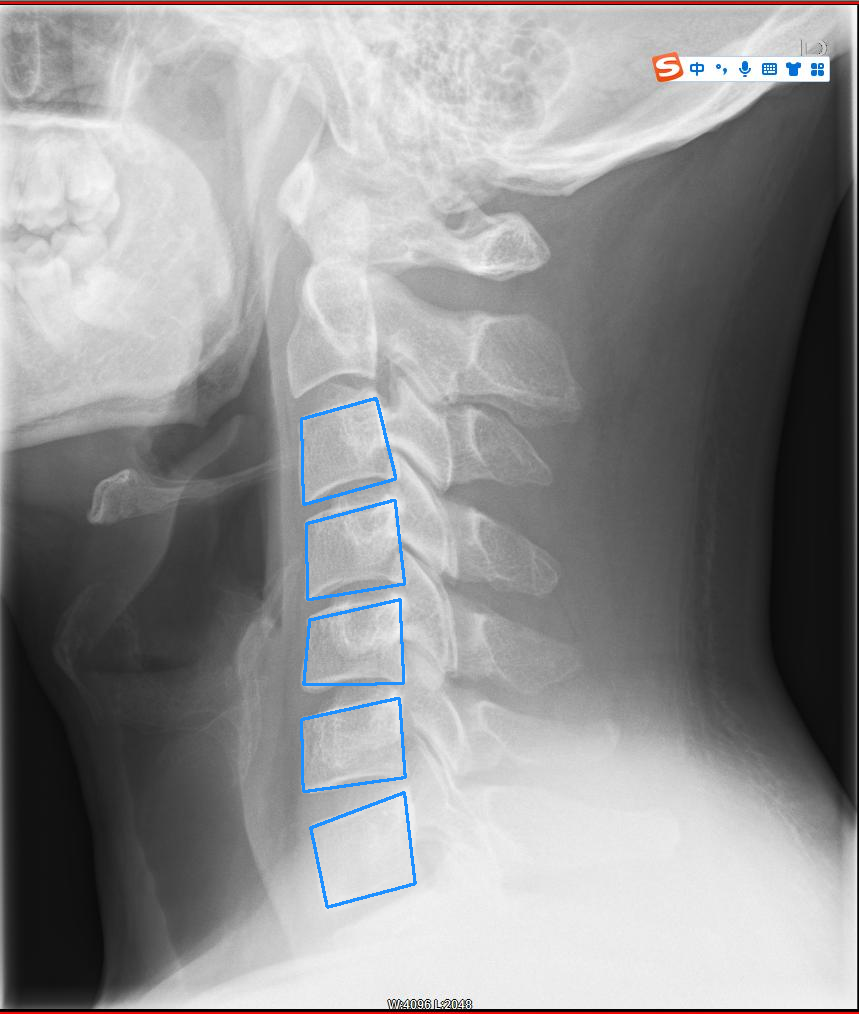}
        \label{fig:sub3}
    \end{subfigure}
    \caption{Example X-ray images from NHANES II (left, center) and CSXA (right) datasets, with annotated vertebrae outlined in blue.}
    \label{fig:three_images}
\end{figure}

\begin{figure*}[t]
    \centering
    \includegraphics[width=\textwidth]{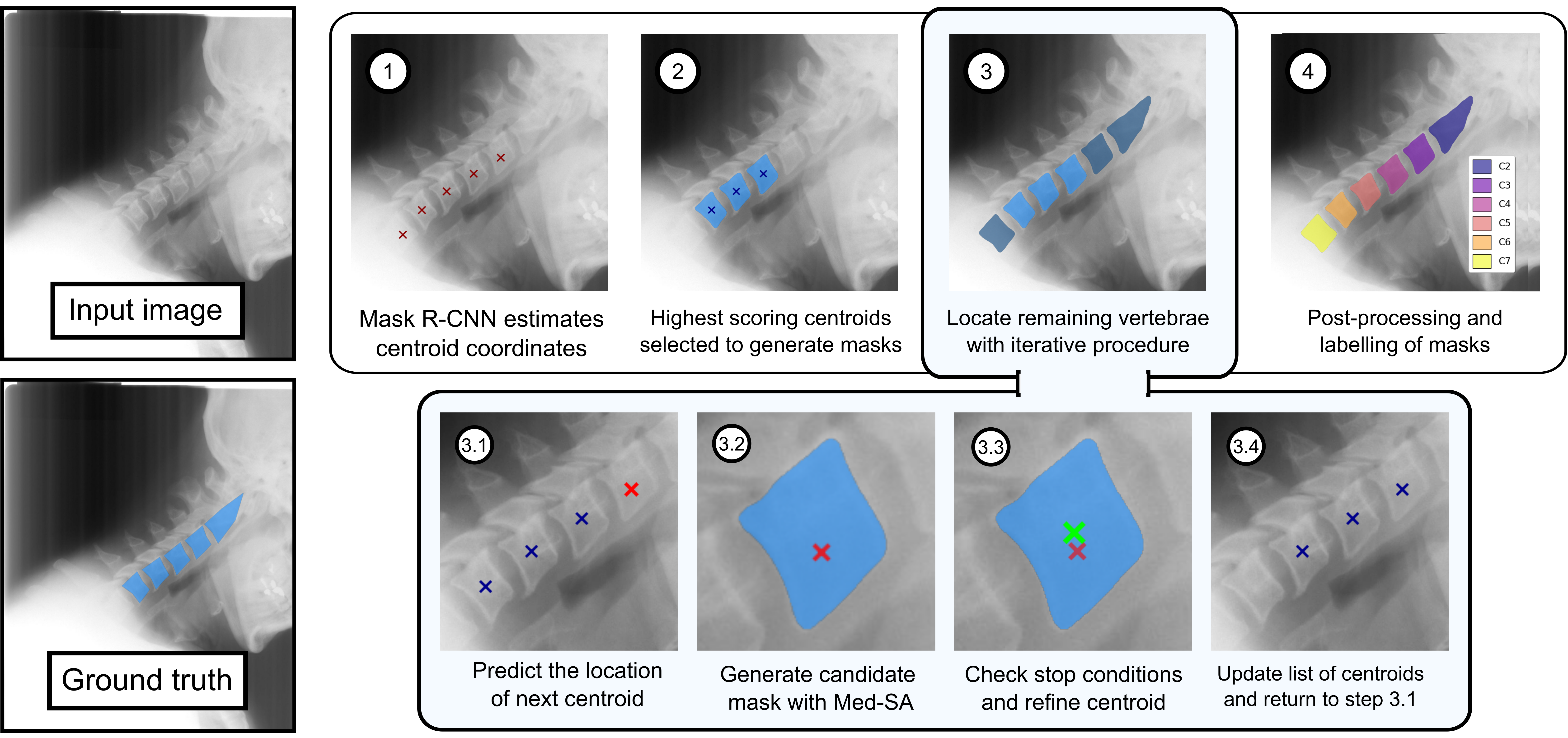}
    \caption{Diagram of the SpineFM pipeline for automated vertebra segmentation in spine X-rays.}
    \label{fig:2}
\end{figure*}

The Segment Anything Model (SAM) \cite{SAM} has demonstrated strong performance in natural image segmentation. However, this success has not fully translated to the medical imaging domain, where images often exhibit lower contrast and ambiguous boundaries \cite{huang_segment_2024}. To address this limitation, the Medical-SAM-Adaptor (Med-SA) was proposed \cite{Med-SA}, which incorporates adapter modules within the image encoder, allowing for fine-tuning while keeping the base SAM weights frozen. During development, we found that this model can effectively segment individual vertebrae, making Med-SA an ideal backbone for our spine segmentation pipeline.

Our contributions are as follows. We introduce SpineFM, a novel framework that leverages Med-SA as a general-purpose segmentation foundation model for precise segmentation of challenging spinal radiographs with limited labels. To initialize our pipeline, we first employ R-CNN to obtain rough estimates of vertebrae locations, and the most confident estimates are then input into Med-SA for refined vertebrae segmentation. Following this step, SpineFM applies an inductive approach to locate neighboring vertebrae by exploiting the spine’s ladder-like anatomical structure. The inductive nature of SpineFM, combined with the efficiency of the segmentation foundation model, enables robust performance across a range of anatomical variations, contrasting with traditional U-Net-based or Region-CNN spine segmentation methods, and compensates for missing manual annotations. Consequently, SpineFM achieves state-of-the-art results for both cervical and lumbar spine sections.

\section{Data}
\label{sec:format}
Our method was developed and tested using two publicly available datasets: the Second National Health and Nutrition Examination Survey (NHANES II) \cite{NHANES-II} and the Cervical Spine X-ray Atlas (CSXA) \cite{CSXA-dataset}. See Fig. \ref{fig:three_images} for example images.

NHANES II includes 17,000 spine X-ray images covering cervical and lumbar regions, with 544 images manually annotated. These annotations provide landmark coordinates for the C2-T1 vertebrae in cervical scans and T12-S1 in lumbar scans, which were used to generate ground truth masks. However, some images lack annotations for certain vertebrae due to limited visibility caused by poor contrast or limited field of view (FOV). We split this dataset into 70\%, 15\%, and 15\% for training, testing, and validation, respectively.

CSXA comprises 4,963 cervical spine X-rays, each with complete annotations marking all four corners of the C3-C7 vertebrae. Unlike NHANES II, CSXA images are of higher quality and exhibit consistent FOV. We used a subset of 600 samples from this dataset, split evenly between training, testing and validation.

\section{Methodology}
\label{sec:pagestyle}

Unlike previous studies \cite{kim_automated_2021,VertXNet}, which located all vertebral bodies simultaneously, we aim to exploit the repeatable structure of the spine by iteratively predicting the centroid of the next vertebra based on the centroid of previous vertebrae. This inductive approach is similar to methods used for predicting vertebra corners from spinal CT scans \cite{ladder}. To initialize this inductive process, our pipeline must first estimate the locations of enough vertebrae to infer the remaining ones. Additionally, we incorporate a stopping mechanism to interrupt the process when necessary. To achieve these functionalities, we combine several distinct models into a multi-step pipeline (illustrated in Fig. \ref{fig:2}), which can be divided into two main stages: initial predictions and inductive predictions, as outlined next.

\subsection{Initial Prediction}
\label{ssec:subhead}
The main goal of the initial stage is to predict the centroid coordinates of three successive vertebrae from the original image. The first step involves using a region-based CNN model, Mask R-CNN \cite{Mask-RCNN}, to generate a list of potential vertebrae masks along with corresponding confidence scores. Masks with confidence scores below 0.6 are immediately discarded. The remaining masks are then used to calculate centroid coordinates, which are sorted based on their projection onto a line of best fit through all points. The three coordinates with the highest average confidence scores are selected as the initial estimates. A patch of the X-ray around each coordinate is fed into the Med-SA, which generates the first three masks. These masks are subsequently used to recalculate the estimated centroid coordinates, forming the basis for the inductive stage.
Unlike previous approaches \cite{ladder,VertXNet}, our method does not require the detection of a predefined reference vertebral body to initialise segmentation. This characteristic of SpineFM enhances robustness, particularly when these vertebrae are outside the X-ray's field of view.

\subsection{Inductive Prediction}
\label{ssec:subhead}
With the first three centroids estimated, a small fully connected neural network is employed to estimate the coordinates of the next vertebra. A fixed-size patch of the X-ray around this coordinate, approximately twice the size of a vertebra, is fed into the Med-SA model to predict the corresponding mask, which is then used to refine the estimate for the vertebra centroid. Next, we check for overlap between the new mask and the previous vertebra mask by calculating their intersection-over-union (IoU). If the IoU exceeds the threshold of 0.1, the new mask is rejected and the algorithm terminates. If the IoU condition is satisfied, a patch around the new centroid is input into a ResNet classifier, which distinguishes between background, a regular vertebra, or one of the two spine-end vertebrae, S1 and C2.
If the background is detected, the new mask is rejected, and the algorithm terminates. If one of the spine-end vertebrae is detected, the new mask is retained and marked as the spine-end vertebra, terminating the algorithm once again. Finally, if a regular vertebra is identified, the new mask is kept and the process repeats with a new set of three initial points: the new centroid and the two previous centroids. This iterative process continues both upward and downward along the spine.
Our framework is highly data-efficient, as it operates on a patch level, unlike \cite{ladder,VertXNet}, which utilizes the entire image for prediction. This patch-based approach enables to achieve reliable results even with extremely limited training labeled data.

\begin{table}[!ht]
    \centering
    \begin{minipage}{0.48\textwidth}
        \centering
        \includegraphics[width=\textwidth]{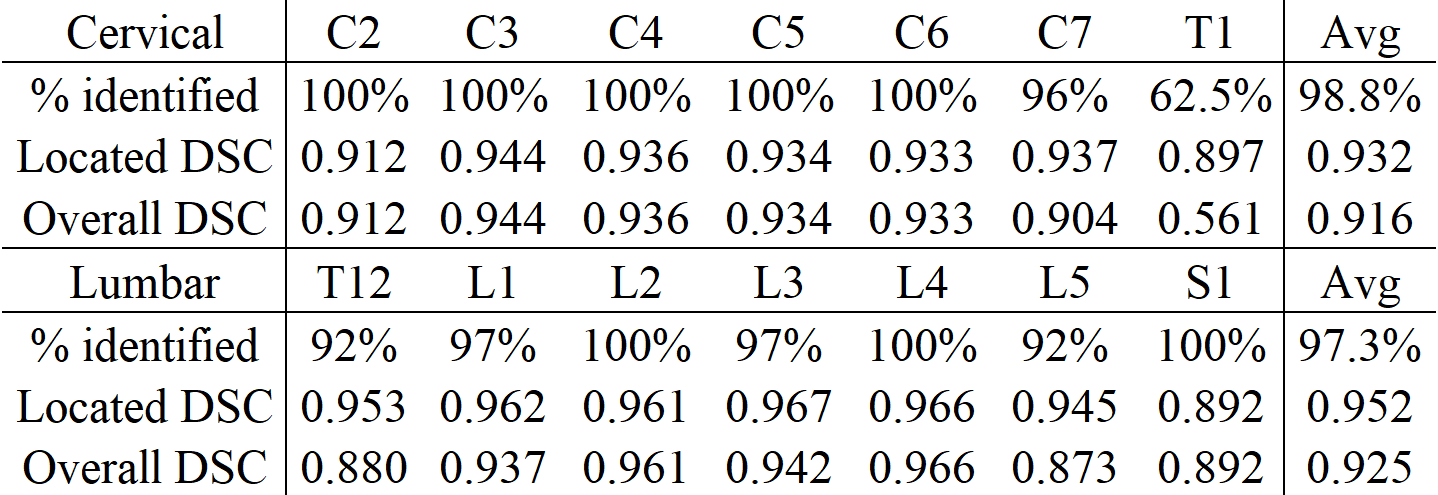}
        \caption{Results for the NHANES II dataset}
        \label{table:1}
    \end{minipage}
    \hspace{0.02\textwidth} 
    \begin{minipage}{0.48\textwidth}
        \centering
        \includegraphics[width=\textwidth]{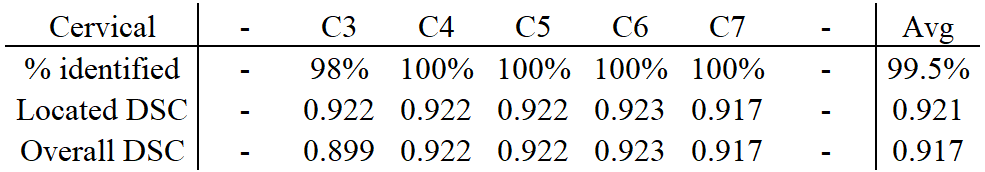}
        \caption{Results for the CSXA dataset}
        \label{table:2}
    \end{minipage}
\end{table}

\subsection{Post-Processing}
\label{ssec:subhead}
Once the final list of masks is generated, each mask is converted from logits to a binary mask using a sigmoid function followed by a threshold of 0.9. Each mask is then smoothed by a Gaussian filter and re-thresholded. The positions of the known spine-end vertebrae are subsequently used to infer the labels of the remaining vertebrae along the spine.

\subsection{Models}
\label{sec:typestyle}
This section provides brief details on each model used within the pipeline. Further details on specific training the parameters for each model can be found in the supplementary code.

\subsubsection{Medical-SAM-Adaptor}
\label{sssec:subsubhead}
The Med-SA \cite{Med-SA} model is employed to segment individual vertebrae when provided with an initial estimation of the centroid coordinates. During training, samples are generated by selecting a random point within the ground truth mask and creating a patch around that point. For each epoch, the model generates a new patch for each ground truth mask. The large variation in prompts ensures that the model remains robust to inaccuracies in centroid coordinates during inference.
It is important to note that the publicly available version of Med-SA includes modifications only to the SAM image encoder, while the prompt embedder and mask decoder remain unchanged.

\subsubsection{Vetrebra Detection and Classification}
Mask R-CNN is employed for the initial centroid predictions. Beginning with the COCO V1 pre-trained weights, we fine-tuned the model independently on each dataset ie. NHANES II and CSXA.
We utilize ResNet-101 \cite{ResNet} to classify predictions as either a spine-end vertebra (S1 or C2), a regular vertebra, or background. We began with weights pre-trained on ImageNet 1K, which were then fine-tuned on our datasets.


\subsubsection{Point Predictor}
\label{sssec:subsubhead}
Our point predictor is a shallow fully connected neural network. It consists of six input channels, representing the three pairs of coordinates for the known centroids. The network features a hidden layer with 50 units and ReLU activation, and it produces two output channels for the next centroid coordinates.

\section{Results}
\label{sec:format}

Tab. \ref{table:1} and Tab. \ref{table:2} present our method's results on each dataset. When evaluating the performance of our method, we individually assessed each vertebra mask prediction. Then, for each vertebra level, we calculated '\% identified'; the percentage of vertebrae from this level that were detected by our model, 'Located DSC'; the average dice similarity coefficient (DSC) for the vertebrae detected by our model, and 'Overall DSC'; the average DSC across all annotated vertebrae, defaulting to zero for each unidentified vertebra. We also included the average of these metrics across each section, accounting for the frequency of occurrence of each vertebra.

\begin{figure}[t] 
    \centering
    \begin{subfigure}{0.48\linewidth}
        \centering
        \includegraphics[width=\linewidth]{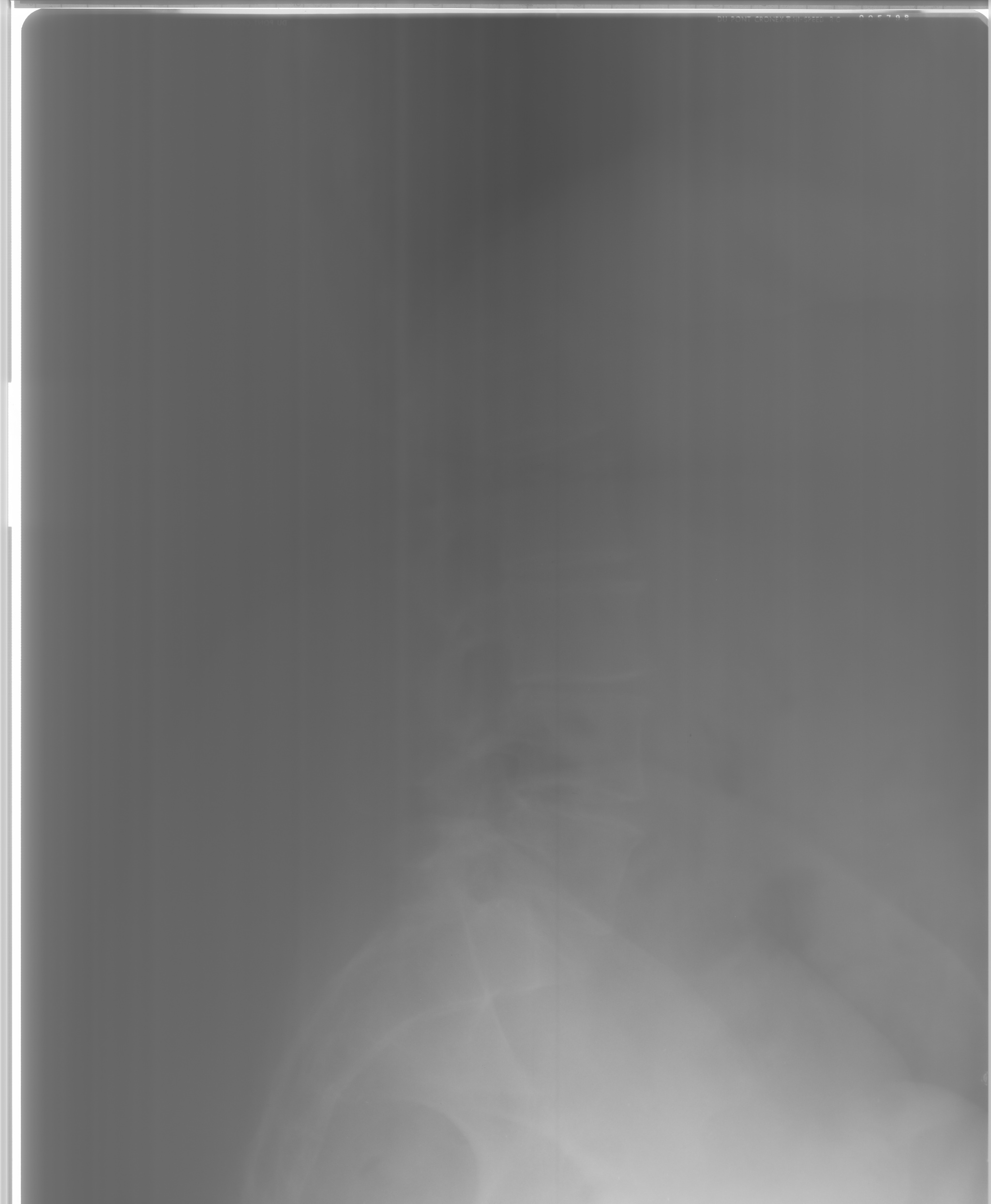} 
        \caption{Low-contrast NHANES II image with reduced vertebrae clarity} 
    \end{subfigure}%
    \hfill 
    \begin{subfigure}{0.48\linewidth}
        \centering
        \includegraphics[width=\linewidth]{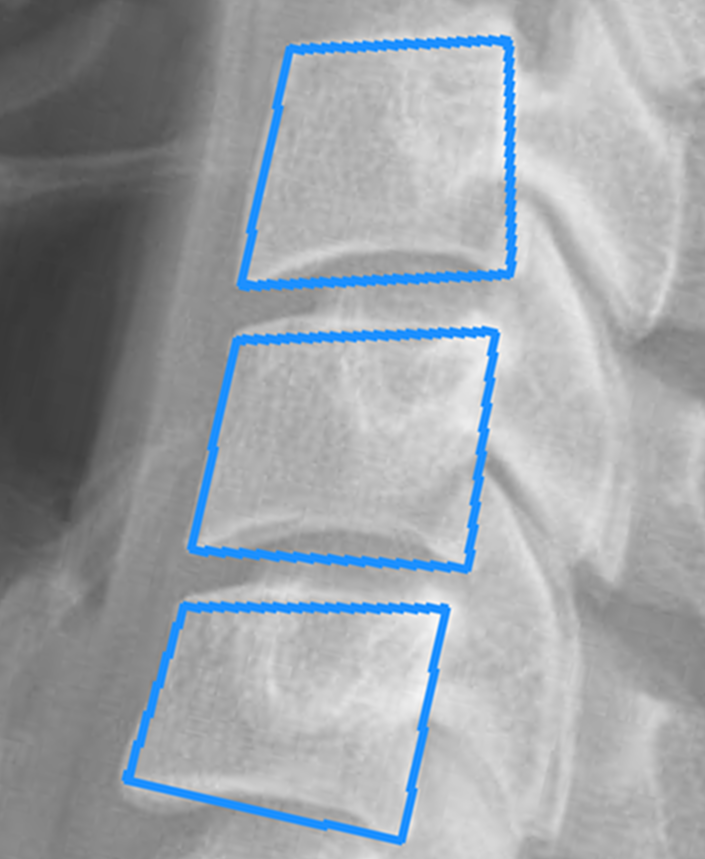} 
        \caption{Poor-quality ground truth masks (e.g. squares) from CSXA dataset}
    \end{subfigure}
    \caption{Examples of limitations related to each dataset, showcasing specific challenges encountered in spine X-ray segmentation} 
    \label{fig:combined} 
\end{figure}

\section{Discussion}
\label{sec:format}
In this paper, we introduced SpineFM, a novel method for spine X-ray segmentation. This method employs an inductive approach to localize each vertebra and utilizes a foundational model to produce high-quality masks. SpineFM significantly outperforms the existing state-of-the-art method, VertXNet, on the NHANES II dataset, achieving overall DSCs of 0.916 and 0.925 for the cervical and lumbar sections, respectively, compared to VertXNet’s DSCs of 0.880 and 0.863 \cite{VertXNet}.
Mask R-CNN and nnUNet, two widely used models, have also been assessed for spine segmentation \cite{VertXNet}. On cervical and lumbar samples from an in-house dataset, Mask R-CNN achieved average DSCs of 0.730 and 0.649, while nnUNet achieved average scores of 0.903 and 0.829. Both are significantly surpassed by SpineFM's performance for NHANES II.
Finally, MDR2U-Net, another state-of-the-art approach that segments the L1-L5 vertebrae, achieves an average DSC of 0.929 on an in-house dataset \cite{kim_automated_2021}. In contrast, SpineFM achieves an average DSC of 0.936 for the same set of vertebrae on the NHANES II dataset.
Unfortunately, since we are the first study to use the CSXA dataset, we do not have other examples to compare our results to. 


Tab \ref{table:1} demonstrates that the robustness of our pipeline, achieving an average DSC of over 0.93 across most vertebra levels, with some levels even surpassing 0.96. Achieving such high DSCs is challenging due to inherent ambiguities in defining the exact boundaries within X-ray images. Furthermore, since each ground truth mask is constructed as a polygon from just seven landmark vertices, and so it may not perfectly represent the true shape of the corresponding vertebra (see Fig. \ref{fig:combined}(b)).

The main limitation of our method in the NHANES II test arose from imperfect vertebra identification, made apparent in the discrepancy between the Located DSC and Overall DSC. This issue occurs when the patch classifier misclassifies a new vertebra mask, leading to its rejection and the termination of the iterative process. Vertebrae at the ends of the spinal sections were particularly affected, especially T1 in the cervical section and T12 and L1 in the lumbar section. This misclassification is primarily due to inconsistent saturation levels between the X-rays, which can result in very poor contrast (see Fig. \ref{fig:combined}(a)).
It is important to note that the NHANES II X-rays are digital scans of physical X-ray films. In contrast, digital X-rays offer superior clarity and contrast, which would mitigate this issue in a clinical setting. Our results on the CSXA dataset, which contains digital X-ray images, support this conclusion, as we successfully identified 99.5\% of all labeled vertebrae.

While all CSXA X-ray images are of high quality and have consistent FOV, we observed a similar overall DSC to the NHANES II dataset, along with a decrease in the average DSC for the located vertebrae. We believe that this decrease is not due to a decline in segmentation performance but rather a consequence of the ground truth masks, which inadequately evaluate segmentation quality. Since the dataset only provides annotations for the vertebra corners, each mask is represented as a simple quadrilateral that does not accurately reflect the true shape of the underlying vertebrae. This limitation is evident in Fig.\ref{fig:combined}(b). Consequently, the DSCs calculated from these masks do not truly represent the quality of our predictions.

\section{Conclusion}
\label{sec:format}
This paper presents SpineFM, a novel method for vertebral body segmentation that employs an innovative inductive approach for locating each vertebra and utilizes a powerful foundational model for its effective segmentation. We demonstrate that SpineFM achieves state-of-the-art performance on two separate datasets, NHANES II and CSXA, which include both cervical and lumbar spine sections.
Future work may involve employing an agent-based mechanism to model our inductive process of vertebrae identification as state evolution in a dynamic process \cite{mo2024labelling} to further automate our pipeline.

\section{Acknowledgments}
\label{sec:acknowledgments}
This research study was conducted retrospectively using human subject data made available in open access \cite{CSXA-dataset,NHANES-II}. Ethical approval was not required as confirmed by the license attached with the open-access data. The project was funded by Oxford Big Data Summer Internship Programme 2024. The model weights and code can be accessed here: https://github.com/sjsimons/SpineFM

\bibliographystyle{IEEEbib}
\bibliography{strings,refs}

\end{document}